\documentclass[12pt,preprint]{aastex}

\newcommand{\msun} {$M_\odot$}
\shorttitle{Proper Motion Objects in the HDF}
\shortauthors{Kilic et al.}

\begin{document}

\title{Proper Motion Objects in the Hubble Deep Field\footnote{Based on observations made with the NASA/ESA Hubble Space Telescope, obtained from the Data Archive at the Space Telescope Science Institute, which is operated by the Association of Universities for Research in Astronomy, Inc., under NASA contract NAS 5-26555.}}

\author{M. Kilic, Ted von Hippel}
\affil{The University of Texas at Austin, Department of Astronomy, 1 University Station 
C1400, Austin, TX 78712, USA}
\email{kilic@astro.as.utexas.edu, ted@astro.as.utexas.edu}

\and

\author{R. A. Mendez}
\affil{European Southern Observatory, Casilla 19001, Santiago, Chile}
\email{rmendez@eso.org}

\and

\author{D. E. Winget}
\affil{The University of Texas at Austin, Department of Astronomy, 1 University Station
C1400, Austin, TX 78712, USA}
\email{dew@astro.as.utexas.edu}

\begin{abstract}
Using the deepest and finest resolution images of the Universe acquired with the Hubble Space
Telescope and a similar image taken 7 years later for the Great Observatories
Origins Deep Survey, we have derived
proper motions for the point sources in the Hubble Deep Field--North. 
Two faint blue objects,
HDF2234 and HDF3072, are found to display significant proper motion, 10.0 $\pm$ 2.5 and 15.5
$\pm$ 3.8 mas yr$^{-1}$. Photometric distances and tangential
velocities for these stars are consistent with disk white dwarfs located at
$\sim$ 500 pc.
The faint blue objects analyzed by Ibata et al. (1999) and Mendez \& Minniti (2000) do not
show any significant proper motion; they are not halo white dwarfs and they do not
contribute to the Galactic dark matter. These objects are likely to be distant AGN.
\end{abstract}


\section{Introduction}

Major observational campaigns have searched for dark matter in the form of massive compact halo objects 
(MACHOs) using microlensing events (e.g. Alcock et al 1997; Afonso et al. 2003; Udalski et al. 
1992). The detection of 13--17
microlensing events toward the Large Magellanic Cloud during 6 years by the MACHO 
collaboration implies that a significant fraction (20$\%$) of the halo of 
the Galaxy may be in the form of compact halo objects (Alcock et al. 2000).
The time scale of these lensing events eliminates the possibility of MACHOs having 
substellar masses. The MACHO collaboration finds a most probable mass of 0.5 \msun\,
which supports the idea of a massive halo comprised of baryonic matter in the form 
of low luminosity white dwarfs (Kawaler 1996).
Recent observations by the EROS group provide further evidence that
less than 25\% of a standard dark matter halo can be composed of objects with a mass between
2 $\times$ 10$^{-7}$ \msun\ and 1 \msun\ (Afonso et al. 2003). 

Halo white dwarf stars are expected to have
large proper motions as a result of their high velocities relative to the Sun.
HST proper motion studies of the Globular
 Cluster NGC 6397 showed that most of the required dark matter in the solar vicinity 
can be accounted for by a population of old white dwarfs representing the thick disk 
and halo of the Galaxy (Mendez 2002).
Claims by Oppenheimer et al. (2001) and Ibata et al. (2000) that they had found a 
significant population of halo white dwarfs from kinematic surveys are tantalizing. 
Their discoveries seemed to be consistent with earlier findings of an old population 
of white dwarfs in the Hubble Deep Field (Mendez \& Minniti 2000). However, further 
analysis by several groups showed that the sample of Oppenheimer et al. (2001) could 
also be interpreted 
as the tail of a kinematically warmer white dwarf component, better explained by the thick disk 
population of the Galaxy (Reid et al. 2001; Reyle et al. 2001; Mendez 2002; 
Bergeron 2003).

The Hubble Deep-Field (HDF) provides a unique window on the Universe (Williams et al. 
1996; Flynn et al. 1996). The extreme depth of the HDF provides an unprecedented 
advantage to find faint stellar objects as well as to study very distant galaxies. 
The advantage of going deep is that it allows us to search for faint stellar components 
of the Galaxy in the regions of the color--magnitude diagram that are devoid of any 
contamination by standard Galactic stars. The lack of ordinary disk stars is due to 
the finiteness of the Galaxy (Flynn et al. 1996). Mendez \& Minnitti (2000) claimed that 
the faint blue objects found in the HDF--North and HDF--South are Galactic stars based on the
observed number of blue sources and extragalactic sources in the two fields. Independent 
proper motion measurements for five of these faint blue sources by Ibata et al. (1999) 
suggested that they are
cool halo white dwarfs which could account for the entire missing mass in the solar
neighborhood. Third epoch data on these five objects, however, did not show any significant
proper motion
(R. Ibata, private communication; Richer 2001). 

We use the original Hubble Deep Field -- North data and images of the same field taken 
7 years 
later for the Great Observatories Origin Deep Survey (GOODS) to measure proper motions 
of the point sources analyzed by Ibata et al. (1999) and Mendez \& Minniti (2000).

\section{Proper Motion Measurements}

GOODS is a multi-wavelength, multi-facility deep survey designed to study galaxy 
formation and evolution over a large redshift range. It includes deep imaging with 
ACS in the B, V, i, and z bands, and reaches down to AB $=$ 28.1, 28.4, 27.7, and 27.6 in the 
four bands, respectively (10 sigma, point source)
(Giavalisco et al 2003).
Our second epoch data, acquired
with HST and ACS as part of the GOODS ACS Treasury program, provide a baseline of 7 years.
The GOODS team released version 1.0 of the reduced, calibrated, stacked, and mosaiced images
of the HDF -- North in 17 sections. 
Section 32 (total integrations of 34.9 ks in V and 36.9 ks in I) 
and section 33 (48.9 ks in V and 51.9 ks in I) overlap with the original HDF--North images. 

The source catalogs for the first epoch are produced by the Space Telescope Science
Institute (STScI) from the combined and drizzled images. We note that the first epoch 
HDF--North catalogue
is based on rereduced HDF--North images by Casertano et al. (2000), providing a 10\%
increase in depth.  
We used the SExtractor package (Bertin \& Arnouts 1996), version 2.3, to build source
catalogues from the second epoch data.
The major motivation for using SExtractor was its 
incorporation of weight maps in modulating the source detection thresholds. Source 
detection was carried out on the inverse-variance-weighted sum of the V and I band 
drizzled images. The combined V + I image is deeper than any of the individual images 
(Casertano et al. 2000).
Only those objects matching the positions of the objects in the 
first epoch data with differences less than 0.2 arcsec 
are included in our final catalogue. Furthermore, we visually inspected all of the 
sources used for
our proper motion study to avoid any mismatches.
Although the GOODS Team released version r1.0 of the ACS
multi-band source catalogs, their catalogs are based on z--band detection only
(Giavalisco et al. 2004). Hence
the released catalogs are not appropriate for the study of faint blue objects.
The GOODS data are 0.5 -- 0.8 mag shallower than 
the original HDF
images, therefore we use the first epoch images for photometry. Astrometric and photometric 
data for the point sources in the HDF--North are given in Table 1.
We adopted the calibrated V and I photometry of  
Mendez \& Minniti (2000).  

Although the effective point spread function (ePSF) fitting procedure (Anderson 
\& King 2000) is the most precise astrometric technique for  
HST images, well-exposed star images are required to accurately sample 
the PSF. There are not many
stars in the Hubble Deep Field, and the main source of error in our proper motion 
measurements is the positions of the 
reference compact objects (galaxies). Therefore the ePSF method is not
necessary and was not used for our analysis. 
  
The original HDF images were rereduced and corrected for distortion by Casertano et al. (2000). 
The second epoch data were corrected for distortion
by the GOODS team using the latest (July 2003) coefficients released by the ACS group 
at STScI. Even with these distortion
corrections, however, some distortion remains (Bedin et al. 2003). The effect 
of the
remaining distortion is larger if a global coordinate transformation is used. Instead of performing
a global transformation, we have used the IRAF routine GEOMAP to derive a quadratic 
local transformation for
each star, using a surrounding net of several dozen compact objects (isolated, low
residuals, and not fuzzy).
After mapping the distortions with the GEOMAP package, object coordinates were transformed
to the second epoch positions with the GEOXYTRAN routine.

Figure 1 shows the contour maps for the two bright stars HDF2272 and HDF3072. 
The immediate field around each object is shown with dashed lines crossing at the first
epoch position. The second epoch position is marked with an asterisk.
This figure shows that SExtractor works very well for bright
compact objects and these two objects are apparently moving.
For faint objects pixel maps are more informative than contour maps. 
Pixel maps for two faint, possibly moving objects are shown in Figure 2.
Solid lines cross at the first epoch position, and the second epoch position is marked with
a box.
Centroiding errors for faint stars are naturally worse, therefore
proper motion errors are larger for the fainter stars.

\section{Results}

Proper motion measurements are mainly affected by distortion mapping and selection of reference objects.
The RMS error of the transformations
are larger than the positional errors of the objects. 
In order to check our distortion solution we have used the GEOMAP package with
different polynomial terms. We started with no distortion correction and deleted 
deviant points using a 3 $\sigma$ 
rejection algorithm. Rejection of very deviant points is required due to the fact that our reference objects are
compact galaxies and centroiding errors are larger for galaxies. We used quadratic, 
quadratic with one cross--term, and quadratic with
4 cross--terms local transformations. For most of the objects, 
the results from higher--order
transformations were very similar to the results from the quadratic (with no
cross--term) local transformation.
This gave us confidence in the stability of our procedure. For two objects,
the use of the higher--order terms made the distortion solution unstable because of the 
relative positioning of the 
reference objects. To be conservative, we adopted the quadratic with no cross--term
local transformation for distortion mapping for all of our 
objects. 

Figure 3 shows the differences between second epoch coordinates and transformed first epoch coordinates
for one of our stars, HDF1583, and the surrounding 40 reference objects.
A 3 $\sigma$ rejection algorithm is later used to eliminate outliers from the sample.
Error bars include positional errors from the SExtractor first
and second epoch coordinates and the RMS error of the transformation. It is clear from 
this figure that 
HDF1583 is statistically well separated from the reference objects,
most or all of which are galaxies:
it is moving with respect to this external reference frame. 

In order to further test our transformations, we have also used all compact objects 
with positional differences between the two epochs of
less than 0.4 pixels to perform a global transformation. We found 377 compact galaxies matching 
our criteria, and fit a quadratic polynomial
to map the distortions. As described above, we have measured proper motions in four to six different ways. Our conservative estimate of
the proper motions, their significance ($\mu/\sigma$), and position angle are given in 
Table 2, along with the observed
range of proper motions from different transformation versions. A comparison of the observed ranges and
errors for the proper motion measurements
show that the errors are consistent with the variations between fitting techniques.
Typical errors in our measurements are $\sim$2.5 mas yr$^{-1}$. Hence, only those objects
having proper motions larger than 5 mas yr$^{-1}$ have significance greater than two. 
The bright objects HDF2272, HDF2234, HDF101, HDF1583, HDF3072, 
HDF2258, and HDF1481 are definitely moving, and the faint objects HDF1816 and HDF774 might 
be moving. 

Star--galaxy confusion becomes worse at faint magnitudes. Only objects 15$\sigma$ above the
sky level were analyzed by Mendez \& Minniti (2000). Proper motions provide further 
star-galaxy separation since anything with a significant proper motion cannot be very
distant (e.g. Ibata et al. 1999). 

We calculated photometric distances for all objects
in our sample assuming that they are either main sequence stars, white dwarfs, or white
dwarfs on the blue hook of the cooling sequence. 
For a given $V-I$ color, we estimate three absolute magnitudes for each object
by linearly interpolating the $V-I$ and $M_{V}$ relation for main sequence stars
(Table 15.7 of Allen's Astrophysical Quantities, 2000), white dwarfs, and cool white 
dwarfs (Hansen et al. 1999).

White dwarfs become redder as they cool until the effects of collision induced
absorption due to molecular hydrogen becomes significant below $\sim$5000 K. 
The $V-I$ colors for white dwarfs are expected to become 
bluer for $T_{\rm eff}\lesssim3500 K$ (Hansen et al. 1999;
Saumon \& Jacobson 1999).
Cool white dwarf colors can be                     
quite different
in Johnson/Kron-Cousins and the HST filters since H$_2$ opacity produces sharp flux
peaks in the white dwarf spectra. The observed colors of the white dwarfs depend on
the transmission peaks of the filters. Richer et al. (2000) calculated the HST colors
for white dwarfs using the Holtzmann et al. (1995) bandpasses and the transformations
they use to express fluxes in V, R and I. Hubble Deep Field photometry is calibrated
using the Holtzmann et al. (1995) transformations (Mendez \& Minniti 2000). Therefore we used Richer et 
al. (2000) white dwarf cooling tracks instead of more recent Chabrier et al. (2000b) models.
We use the apparent magnitudes of the objects and the adopted absolute magnitudes 
to estimate photometric distances. 

Proper motion measurements and derived distances can be used to calculate tangential
velocities using the equation 
\begin{equation}
\mu=\frac{V_{tan}}{4.74 d}
\end{equation}
where $\mu$ is the proper motion in arcsec yr$^{-1}$, $d$ is the distance in parsecs, 
and $V_{tan}$ 
is the tangential velocity in km s$^{-1}$.  
With the assumption that the objects are either main sequence stars, DA white dwarfs, or 
cool white dwarfs, derived distances and tangential velocities are given in Table 3.
The differences between Chabrier et al. (2000b) and Richer et al.
(2000) white dwarf colors are equivalent to absolute magnitude differences of 0 to 0.5 mag. This corresponds to
0--25\% difference in estimated distances and velocities with an average difference of about 10\%.
  
\subsection{Bright Sources ($V\leq27$)}
 Mendez \& Minniti (2000) analyzed sources brighter than V$=$27,
for which SExtractor
gives reliable star--galaxy separation.
The same sources have stellarity indices $\geq$ 0.97 in the GOODS data (Giavalisco et al.
2004) which has better spatial resolution than the original HDF images. The morphology
of these sources as point--like is well supported.
HDF2272, HDF2234, HDF101, HDF1583, HDF1828, HDF2134 and HDF2258 are further
confirmed to be stars with Keck LRIS spectroscopy (Cohen et al. 2000). 

We have searched the 503 X--ray point sources detected
in the 2 Ms Chandra exposure of the region around the Hubble Deep Field North called
the Chandra Deep Field North (Barger et al. 2003) for possible matches with
the point sources analyzed here.
We did not find any objects matching our objects within a search radius of 0.5 arcseconds;
we could not confirm if we had any quasars among our objects.

A comparison of the distances, tangential
velocities, and photometric colors show that HDF1828, HDF2134, HDF2258, HDF1481, HDF684,
and HDF3000 are halo main sequence stars (Table 3). Their $V-I$ colors are too red to 
be white dwarfs. 
HDF2272, HDF101, HDF1583, and HDF1470 can be either main sequence stars or 
white dwarfs. 
Since we are sampling a larger volume for main sequence stars,
these four stars are likely to be main sequence stars.
Cohen et al. (2000) classified HDF2272 and HDF1583 as stars showing Mg absorption and
Balmer lines, and HDF2234, HDF101, HDF1828, HDF2134 and HDF2258 as stars showing TiO
or CaH bands.

 HDF2234, HDF3072, HDF161, HDF3031, and HDF759 would have to be at very large distances 
and moving with velocities higher than the escape velocity of our Galaxy if they were main 
sequence stars.
The first 11 objects (V $\leq$ 26) in Table 1 are also detected in the Hawaii-HDF-N Survey (Capak et al. 2003),
an intensive multi-color (U, B, V, R, I, z, HK) imaging survey of 0.2 square
degrees centered on the HDF-N. Figure 4 shows normalized $UBVRIz$ magnitudes
for HDF2234 (filled triangles), HDF3072 (filled circles), and HDF161 (open circles) along
with colors for a 15000 and 3000 K blackbody (long--dashed lines). The observed magnitudes
are normalized at $V$. Dashed--dotted
lines represent two DA white dwarf models (7000 K and 3500 K), and dotted lines represent
two DB white dwarf models (8000 K and 3500 K; D. Saumon, private communication). Our
simulations
for a QSO at z=0.3 (upper solid line) and another at z=0.7 (lower solid line) are also shown.
We have used the composite quasar spectra from the Sloan Digital Sky
Survey (Vanden Berk et al. 2001) to simulate the colors for quasars. Plotting all of
the normalized magnitudes in the same plot is an efficient way of presenting all of the data;
it is similar to plotting low--resolution spectroscopy.

A comparison of the observed colors of HDF2234 with the models shows that it is hotter than 10000 K.
J. Cohen kindly provided us the Keck/LRIS spectrum for HDF2234. The same object is also observed
by the Team Keck Treasury Redshift Survey (Wirth et al. 2004) who made their data publicly
available. Figure 5 shows the uncalibrated spectrum of HDF2234 observed by the Team Keck
Treasury Redshift Survey.
The object shows H$\alpha$ at its rest wavelength; it is a star in our Galaxy.
Although Cohen et al. (2000) classified this object as a late--type star showing CaH or TiO, its
colors indicate that HDF2234 is too hot to show CaH and/or TiO.
Absence of H$\beta$
and Mg absorption eliminates the possibility of the object being a main sequence star.
The broad feature at $\sim$ 5400 \AA\ might be due to the efficiency of the instrument + blocking
filter combination, and the 6890 \AA\ feature is the atmospheric B band.
The star could be a cool DA white dwarf showing only H$\alpha$.
However, the colors indicate that this should be a white dwarf hotter 
than 10,000 K which is inconsistent with such weak
H lines, unless the star is a DC white dwarf at a position just too cool to show 
He I (11,000K) but then H$\alpha$ alone cannot be explained. The nature of this object remains
ambiguous at this time. Follow-up spectroscopy in the blue is needed to confirm that this object
is a white dwarf. 

HDF3072 displays even higher apparent proper motion than HDF2234,
15.47 $\pm$ 3.83 mas yr$^{-1}$. Its
colors, and implied distance and velocity are consistent with a $\sim$ 4500 K white dwarf at
d $\approx$ 500 pc.
 
The faint blue objects in Mendez \& Minniti (2000) are HDF684, HDF161, HDF3031, HDF759, and HDF995. We have classified HDF684 as a main sequence star (see above). The rest of the faint blue
objects, HDF161, HDF3031, HDF759, and HDF995 do not seem to exhibit any proper motion. 
HDF161 was near the
detection limit of the Hawaii-HDF-N, and other three objects are not detected in the Hawaii-HDF-N. Figure 4 shows that HDF161 (open circles) exhibits a near--infrared excess; it is consistent with being a QSO
under the given photometric uncertainties. Therefore we believe that HDF161, HDF3031 and
HDF759 are probably AGN. These objects cannot be low--mass main sequence stars, brown dwarfs,
or free floating planets due to their blue colors (see Chabrier et al. 2000a).
Also, they cannot be comets or asteroids
in our solar system due to their small proper motions (A. Cochran, private communication).
Due to the large errors in
the distance and velocity for HDF995, its nature is unclear.

\subsection{Faint Sources ($27\geq V \geq29$)}
Star--galaxy separation becomes ambiguous below V$\approx$27. Proper motions can be used
to identify stars fainter than 27th magnitude since a moving object has to be
in our Galaxy. We find that only two of the objects in our sample, HDF1816 and HDF774,
have significant movement. These two objects are most likely Galactic white dwarfs.
The rest of the faint objects do not show any significant movement ($\mu/\sigma\leq2$). 
For these objects, distances
and velocities are consistent with halo white dwarfs or extragalactic sources.
Mendez \& Minniti (2000) found 566 extragalactic sources in the same magnitude and color
range as the 5 faint blue sources that are brighter than 27th magnitude. The ratio of the
number of extragalactic objects to the number of stars increases at fainter magnitudes.
Therefore, we believe that faint sources, with no significant apparent proper motion, are
extragalactic objects.

\subsection{Ibata et al. (2000) Objects}    
Ibata et al. (1999) obtained second-epoch exposures of the HDF in 1997, and derived proper
motions using a 2-year baseline. They found that two blue, faint objects displayed proper 
motions $\sim$ 25 mas yr$^{-1}$ and three other stars at the detection limit of the second--epoch observations might be moving. Third epoch data on these
objects showed that these objects are not moving (R. Ibata, private communication; Richer 2001). 

Two of the objects in the Ibata et al. (1999) sample are in common with 
Mendez \& Minniti (2000)
objects. These two objects, HDF806 and HDF1816, have stellarities larger than 0.9, 
therefore are classified as 
stars by SExtractor. Stellarity is the probability of an object being a point
source (stellarity$=$1) or an extended object (stellarity$=$0) assigned by
SExtractor. Mendez \& Minniti (2002) found that all objects with stellarity
$<$ 0.85 are clearly extended, and used a conservative cut at stellarity $>$
0.90 to identify point sources. 
We find that HDF806 and HDF1816 have proper motions of 1.15 
$\pm$2.91 mas yr$^{-1}$
and 5.49 $\pm$1.89 mas yr$^{-1}$, respectively.
The other three objects are classified as galaxies by the SExtractor. Visual inspection of 
the first and second epoch images (Figure 6) shows that these three objects are extended, and clearly not stars. 
We conclude that 3 of the objects (2-766, 4-141, 4-551) in the Ibata et al. (1999) sample are 
galaxies, HDF806 (2-455) is not moving, and HDF1816 (4-492) is probably moving (2.9 $\sigma$ 
significance).

\section{Discussion}
The nature of the faint blue objects in the Hubble Deep Field may be crucial to understanding
the contribution of low luminosity halo white dwarfs to micro-lensing events and the dark matter 
content of the Galaxy. Apparent proper motions for 5 faint blue objects (Ibata et al. 
1999) was enough to explain the entire missing mass in the halo of the Milky Way. 
Mendez \& Minniti (2000) claimed that the faint blue objects are white dwarf stars 
located at heliocentric distances of up to 2 kpc and belong
to the Galactic halo. They found a local
halo white dwarf mass density of 4.64 $\times$ 10$^{-3}$ \msun\ pc$^{-3}$, which would account
for about 30--50$\%$ of the dark matter in the Galaxy. 

With the advantage of a 7--year baseline, we are able to place better limits on the proper
motion measurements of the faint blue objects. Using the proper motion information, we also
derived distances and tangential velocities for these objects. Figure 7 shows the tangential
velocities and distances for objects brighter than $V\approx27$ assuming that they are main
sequence stars or DA white dwarfs.
All of the main sequence stars exhibit halo kinematics and distances, whereas all of the likely 
white dwarfs exhibit disk kinematics and distances. 

Following Gilmore, King, \& van der Kruit (1989; see also von Hippel \& Bothun 1990) 
we use the analytical form of the density profile for the thin disk and thick disk 
\begin{equation}
\frac{\nu_{0}(z)}{\nu_{0}(0)}= {0.96}\ {e^{-z/250 pc}} + {0.04}\ {e^{-z/1000 pc}}
\end{equation}
with a local normalization of 0.11 \msun\ pc$^{-3}$ (Pham 1997).
We use the form
\begin{equation}
\nu_{halo}(r) \propto \frac{exp[-7.669\ (R/R_{e})^{(1/4)}]}{(R/R_{e})^{(7/8)}}
\end{equation}
for the halo (Young 1976), where $R$ is the distance from the Galactic center, 
and R$_{e}$ is the scale factor.
$R$ is related
to the distance $r$ from the observer to a star by
\begin{equation}
R^{2} = R_{0}^{2} + r^{2} - 2 r R_{0}\ {\rm cos}b\ {\rm cos}l
\end{equation}
with $R_{0}$ the solar Galactocentric distance,
and {\it b} and {\it l} the Galactic
coordinates for the HDF--North. We use R$_{0}=$7.8 kpc (Gilmore, King, \& van der Kruit 1989), 
R$_{e}=$2.7 kpc (de Vaucouleurs \& Pence 1978) and a local
normalization for the halo of (1/800) $\times$ 0.11 \msun\ pc$^{-3}$ 
(Chen et al. 2001; Gilmore, King, \& van der Kruit 1989).
Using equations 2, 3, and 4, we calculated the expected number of stars in the HDF.
We expect to find 2 thin disk, 3 thick disk, and 11 halo objects in the HDF--North. 
 
We have also used Reid \& Majewski (1993) star count models to predict the number of 
stars in the HDF--North. We found that 2 thin disk, 4 thick disk, and 14 halo objects are expected
in the HDF--North. 
Both simple analytical models and more sophisticated star count models,
when extrapolated to the photometric depth of the HDF,  
predict similar number of stars (16--20) in the HDF--North.

There are 14 stars brighter than $V=27$ and 17 objects fainter than $V=27$
classified as stars by SExtractor.  
The observed number of stars and the predictions of star count models are in good agreement
for $V\la27$   
(see also Mendez et al. 1996 and Mendez \& Minniti 2000). On the other hand, there seems to be
an excess of point sources in the Hubble Deep Field -- North for $V\ga27$. Unfortunately, SExtractor
classification cannot be trusted at these magnitudes. Furthermore, we did not detect
significant proper motion for all but two of these objects.
The two faint, possibly moving objects, HDF774 and HDF1816, may be halo white dwarfs.
One of the problems with any analysis using these objects is that the observations are beyond the
completeness limit, and any calculation based on them is subject to a significant completeness correction. The rest of the objects fainter than $V=27$ are probably extragalactic objects (see
section 3.2).

The five faint blue objects analyzed by Mendez \& Minniti (2000) do not exhibit any significant
proper motion; they are not halo white dwarfs. These objects do not account for the MACHO optical depth and are not the source of the Galactic dark matter.
Their stellar nature is not confirmed either.
The colors of HDF161 are consistent with our QSO simulations. The faint blue
objects may be distant AGN.

Holberg et al. (2002) used a local sample of white dwarfs complete out to 13 pc, 
and found
the local mass density of white dwarf stars to be 3.4 $\pm$0.5 $\times$ 10$^{-3}$ 
\msun\ pc$^{-3}$. Using this normalization factor in equations 2 and 3, we estimate the
expected number of white dwarfs in the Hubble Deep Field. We expect to find 0.05 disk
white dwarfs, 0.09 thick disk white dwarfs, and 0.33 halo white dwarfs in the Hubble Deep
Field North. We have also used Reid \& Majewski (1993) star count models to predict the
number of white dwarfs in the HDF. The results are roughly consistent: 0.10 disk, 0.25 thick disk, and 0.5 halo white dwarfs
are expected. 
  
We have discovered two likely white dwarfs, HDF2234 and HDF3072,
brighter than $V=27$ in the HDF--North. They are located at distances of $\sim$ 500 pc
and have tangential velocities 
$\sim$30 km s$^{-1}$. Their kinematic properties are consistent with being thin disk or 
thick disk objects (see Table 3 and Figure 7).
The expected number of thin disk + thick disk white dwarfs is found to be 0.14 -- 0.35.
We have found 6 to 14 times more disk white dwarfs in the HDF--N than expected from the models.
Assuming Poisson statistics, the probability of finding two white dwarfs is 1\% if the expected number of white dwarfs is 0.14 and
4\% if the expected number of white dwarfs is 0.35. The number of disk + thick disk white dwarfs
may be substantially underestimated. Due to small number statistics, however, this statement is only
a 2--3 $\sigma$ result and it heavily depends on the fact that HDF2234 and HDF3072 are white dwarfs.
Follow--up spectroscopy of these two objects is needed to confirm this result.

Mendez \& Minniti (2000) have found 22 Galactic stars and 10 faint blue objects in the Hubble Deep Field -- South. A
natural test to check the space density of disk and halo white dwarfs would be to obtain second
epoch observations of the HDF--South to find high proper motion objects.
Also, the HST/ACS Ultra--Deep Field observations of the Chandra Deep Field -- South
will be useful to search for faint blue objects at fainter magnitudes and to improve
the morphological classification of these objects at brighter magnitudes.
The Ultra--Deep Field will be $\sim$ 1.5 mag deeper than the HDF and HDF--South (Beckwith et al. 2003).

\acknowledgements
We thank Judy Cohen for kindly providing us Keck/LRIS spectrum of HDF2234. We also thank
Didier Saumon for making his cool white dwarf models available to us and to the Team Keck Treasury
Redshift Survey for making their data publicly available. We are grateful to J. Liebert \& A. Cochran for
useful discussions on the nature of HDF2234. 
This material is based upon work supported by the National Science Foundation under Grant No. 0307315.

\clearpage
\begin{deluxetable}{lrrrrrr}
\tabletypesize{\footnotesize}
\tablecolumns{7}
\tablewidth{0pt}
\tablecaption{Point Sources in the Hubble Deep Field}
\tablehead{
\colhead{Object}&
\colhead{X(HDF)\tablenotemark{a}}&
\colhead{Y(HDF)\tablenotemark{a}}&
\colhead{X(GOODS)\tablenotemark{b}}&
\colhead{Y(GOODS)\tablenotemark{b}}&
\colhead{$V$}&
\colhead{$V-I$}}
\startdata
HDF2272&869.387&989.094&3952.627&838.889&19.78&1.04\\
HDF2234$^{\rm c}$& 2903.711& 1129.655& 6500.462& 8114.658&20.78&0.20\\
HDF101& 315.227& 3788.366& 4778.925& 4538.35&21.45&1.12\\
HDF1583& 1026.935& 1803.079& 4580.210& 1742.69&22.19&1.47\\
HDF1828& 2448.822& 1573.769& 6184.354& 704.792&24.30&2.38\\
HDF3072$^{\rm c}$& 1877.251& 349.952& 4837.550& 7714.658&24.27&1.29\\
HDF2134& 1194.967& 1255.044& 4489.335& 987.877&24.74&2.52\\
HDF2258& 247.806& 1117.993& 3264.773& 1328.88&24.95&2.91\\
HDF1470& 2440.850& 1935.845& 6368.745& 1148.24&25.33&1.50\\
HDF1481& 3424.519& 1920.592& 7554.512& 601.04&25.87&2.22\\
HDF161& 1388.682& 3818.963& 6098.469& 4001.02&25.74&0.77\\
HDF684& 956.369& 2845.424& 5052.038& 3049.25&26.70&1.82\\
HDF3031$^{\rm c}$& 3380.128& 403.175& 6688.739& 6975.563&26.50&0.44\\
HDF759& 534.482& 2774.917& 4501.172& 3190.23&26.66&0.46\\
HDF3000$^{\rm a}$& 1306.869& 420.559& 4178.915& 8105.979&27.36&3.19\\
HDF995& 642.304& 2475.111& 4471.563& 2767.75&27.08&0.52\\
HDF1022& 401.701& 2504.499& 4195.475& 2932.52&27.87&0.29\\
HDF861& 537.648& 2643.861& 4434.536& 3028.95&27.87&0.06\\
HDF1705& 2113.471& 1690.358& 5839.241& 1025.76&28.11&0.46\\
HDF2729& 1746.126& 710.752& 4866.870& 31.945&28.30&0.70\\
HDF2217& 2766.571& 1181.931& 6359.981& 59.073&28.26&0.34\\
HDF806$^{\rm d}$& 1171.947& 2699.015& 5235.329& 2756.72&28.57&0.61\\
HDF1135& 1064.710& 2306.869& 4895.408& 2337.47&28.78&1.16\\
HDF2991& 291.521& 436.858& 2951.714& 479.141&28.58&0.24\\
HDF1288& 926.550& 2096.186& 4613.698& 2154.15&28.53&--0.07\\
HDF946& 351.213& 2538.371& 4151.918& 2999.19&28.49&--0.37\\
HDF1196& 1030.595& 2279.031& 4839.458& 2321.74&28.79&0.74\\
HDF723& 938.652& 2806.484& 5008.853& 3012.2&28.92&0.94\\
HDF1816$^{\rm d}$& 3006.002& 1601.085& 6874.624& 438.659&29.02&0.45\\
HDF1039& 593.115& 2432.467& 4388.899& 2741.84&28.43&--1.35\\
HDF774& 445.116& 2747.477& 4378.614& 3204.48&28.82&0.03\\
\enddata
\tablenotetext{a}{Chip coordinates from the HDF mosaics}
\tablenotetext{b}{Chip coordinates from GOODS section 33}
\tablenotetext{c}{Chip coordinates from GOODS section 32}
\tablenotetext{d}{Object in common with Ibata et al. (1999)}
\tablecomments{There is an offset between our coordinates and the GOODS Teams r1.0 version
of the source catalogues (Giavalisco et al. 2004). The offset is +300 pixels in
X and +200 pixels in Y.}
\end{deluxetable}

\clearpage
\begin{deluxetable}{lrrcrr}
\tabletypesize{\footnotesize}
\tablecolumns{6}
\tablewidth{0pt}
\tablecaption{Proper Motions}
\tablehead{
\colhead{Object}&
\colhead{$\mu$(mas/yr)}&
\colhead{$\sigma$}&
\colhead{$\mu/\sigma$}&
\colhead{$\mu$(range)}&
\colhead{Pos. Angle}}
\startdata
HDF2272& 8.95&2.48& 3.61& 8.30--10.09&227.0\\
HDF2234& 10.05&2.46& 4.09& 8.67--11.08&298.3\\
HDF101& 6.03&2.50& 2.41& 5.00--6.34&238.0\\
HDF1583& 11.60&2.25& 5.17& 11.48--11.97&215.3\\
HDF1828& 2.47&2.06& 1.20& 1.21--3.08&281.7\\
HDF3072& 15.47&3.83& 4.04& 15.40--17.51&268.0\\
HDF2134& 3.28&2.65& 1.24& 2.84--4.44&184.4\\
HDF2258& 8.34&2.92& 2.86& 8.02--8.86&204.2\\
HDF1470& 4.06&3.06& 1.33& 3.40--4.06&230.6\\
HDF1481& 10.53&2.51& 4.19& 10.32--11.38&156.1\\
HDF161& 1.72&1.53& 1.12& 1.27--2.42&96.5\\
HDF684& 3.71&2.19& 1.69& 3.33--4.42&191.0\\
HDF3031& 1.53&2.68& 0.57& 1.53--3.76&211.3\\
HDF759& 1.37&2.56& 0.53& 0.29--1.73&206.6\\
HDF3000& 3.66&2.35& 1.56& 3.43--5.24&230.7\\
HDF995& 1.24&1.64& 0.76& 1.21--1.44&190.5\\
HDF1022& 3.27&3.06& 1.07& 2.19--4.14&251.7\\
HDF861& 2.22&2.93& 0.76& 1.56--2.22&160.8\\
HDF1705& 3.64&3.00& 1.21& 1.83--5.53&235.1\\
HDF2729& 3.45&1.99& 1.73& 2.88--3.78&106.6\\
HDF2217& 2.01&2.79& 0.72& 1.78--2.86&358.0\\
HDF806& 1.15&2.91& 0.40& 0.88--2.36&90.4\\
HDF1135& 3.45&2.00& 1.73& 3.45--4.11&297.6\\
HDF2991& 3.77&3.28& 1.15& 2.84--3.86&36.6\\
HDF1288& 3.55&2.89& 1.23& 2.17--4.25&206.1\\
HDF723& 2.76&2.42& 1.14& 2.32--3.69&98.4\\
HDF1816& 5.49&1.89& 2.91& 4.77--6.30&129.2\\
HDF1039& 3.13&2.50& 1.25& 2.28--4.44&200.8\\
HDF774& 4.80&2.38& 2.02& 3.56--5.22&251.3\\
\enddata
\end{deluxetable}

\clearpage
\begin{deluxetable}{lrrrrrrrrrrrr}
\tabletypesize{\scriptsize}
\rotate
\tablecolumns{13}
\tablewidth{0pt}
\tablecaption{Photometric Distances and Tangential Velocities\tablenotemark{1,2}}
\tablehead{
\colhead{Object}&
\colhead{d(MS)}&
\colhead{$\sigma_{d}$(MS)}&
\colhead{$V_{tan}$(MS)}&
\colhead{$\sigma_{v}$(MS)}&
\colhead{d(WD)}&
\colhead{$\sigma_{d}$(WD)}&
\colhead{$V_{tan}$(WD)}&
\colhead{$\sigma_{v}$(WD)}&
\colhead{d(CWD)}&
\colhead{$\sigma_{d}$(CWD)}&
\colhead{$V_{tan}$(CWD)}&
\colhead{$\sigma_{v}$(CWD)}}
\startdata
HDF2272& 6194& 141& 263& 73& 100.0& 1.0& 4.24& 1.18& 29.65& 0.04& 1.26& 0.35\\
HDF2234& 61094& 2339& 2911& 720& 554.6& 7.6& 26.43& 6.46& 43.65& 0.02& 2.08& 0.51\\
HDF101& 11803& 240& 338& 140& 189.7& 2.6& 5.42& 2.25& 66.68& 0.24& 1.91& 0.79\\
HDF1583& 10965& 177& 603& 117& 115.3& 2.4& 6.34& 1.23& \nodata& \nodata& \nodata& \nodata\\
HDF1828& 10375& 167& 122& 101& \nodata& \nodata& \nodata& \nodata& \nodata& \nodata& \nodata& \nodata\\
HDF3072& 34674& 630& 2542& 631& 505.8& 9.6& 37.09& 9.20& 264.24& 0.64& 19.37& 4.79\\
HDF2134& 11015& 178& 172& 139& \nodata& \nodata& \nodata& \nodata& \nodata& \nodata& \nodata& \nodata\\
HDF2258& 7482& 174& 296& 104& \nodata& \nodata& \nodata& \nodata& \nodata& \nodata& \nodata& \nodata\\
HDF1470& 44874& 1044& 864& 651& 472.1& 7.6& 9.09& 6.85& \nodata& \nodata& \nodata& \nodata\\
HDF1481& 25119& 553& 1254& 301& \nodata& \nodata& \nodata& \nodata& \nodata& \nodata& \nodata& \nodata\\
HDF161& 214783& 27485& 1748& 1575& 2228.4& 74.6& 18.13& 16.19& 428.55& 1.13& 3.49& 3.11\\
HDF684& 58614& 2236& 1030& 610& \nodata& \nodata& \nodata& \nodata& \nodata& \nodata& \nodata& \nodata\\
HDF3031& 599791& 55449& 4359& 7627& 5128.6& 513.7& 37.27& 65.23& 594.29& 1.65& 4.32& 7.55\\
HDF759& 628058& 58264& 4079& 7635& 5248.1& 566.2& 34.08& 63.83& 639.73& 1.80& 4.15& 7.77\\
HDF3000& 15276& 895& 265& 171& \nodata& \nodata& \nodata& \nodata& \nodata& \nodata& \nodata& \nodata\\
HDF995& 580764& 257760& 3420& 4756& 5728.0& 430.5& 33.73& 44.53& 779.83& 2.26& 4.59& 6.05\\
HDF1022& 1.39e06& 332181& 21556& 20807& 12416.5& 2541.7& 192.56& 184.34& 1132.4& 6.92& 17.56& 16.42\\
HDF861& 2.28e06& 1.65e06& 24031& 36146& 19588.4& 5296.6& 206.46& 277.90& 1153.45& 7.06& 12.16& 16.03\\
HDF1705& 1.22e06& 313129& 21039& 18175& 10232.9& 2995.5& 176.47& 154.46& 1247.38& 11.59& 21.51& 17.74\\
HDF2729& 812831& 302641& 13276& 9129& 8053.8& 1321.2& 131.54& 79.05& 1386.76& 13.07& 22.65& 13.10\\
HDF2217& 1.55e06& 546469& 14754& 21179& 13740.4& 3411.5& 130.79& 184.87& 1348.96& 16.89& 12.84& 17.87\\
HDF806& 1.02e06& 134633& 5564& 14085& 10139.1& 2030.9& 55.30& 140.26& 1555.97& 19.87& 8.49& 21.46\\
HDF1135& 326588& 111650& 5341& 3589& 5128.6& 1655.4& 83.87& 55.56& 1995.26& 151.44& 32.63& 19.04\\
HDF2991& 2.06e06& 780649& 36825& 34953& 18793.2& 6184.8& 335.95& 312.56& 1577.61& 20.18& 28.20& 24.54\\
HDF1288& 4.79e06& 6.39e06& 80608& 125961& 30338.9& 7070.9& 510.55& 432.17& 1570.36& 15.06& 26.43& 21.51\\
HDF723& 508159& 281257& 6652& 6889& 7585.8& 1206.8& 99.30& 88.34& 1923.09& 75.78& 25.17& 22.06\\
HDF1816&1.89e06 &717666 &49213 &25202 &15995.6 &7053.9 &416.51 &232.85 &1896.71 &88.67 &49.39 &17.13 \\
HDF1039&\nodata &\nodata &\nodata &\nodata &\nodata &\nodata &\nodata &\nodata &\nodata &\nodata &\nodata &\nodata \\
HDF774& 3.89e06& 4.61e06& 88595& 113808& 31477.5& 9203.1& 716.90& 412.59& 1786.49& 17.43& 40.69& 20.17\\
\enddata
\tablenotetext{1}{Assuming that the object is either a main sequence star (MS), a white dwarf (WD), or a very cool white dwarf (CWD, T$_{eff}\lesssim3500 K$)}.
\tablenotetext{2}{Distances are in parsecs, and velocities are in km sec$^{-1}$}.
\end{deluxetable}

\clearpage
\figcaption[f1.ps]{Two bright, apparently moving objects. The panels show contour maps and first and
second epoch positions of the stars HDF2272 and HDF3072. The contour maps show the flux 
distribution around each object (20 x 20 pixels, 0.6" x 0.6"). Dashed lines cross at 
the first epoch
position. An asterisk marks the second epoch position.\label{fig1}}

\figcaption[f2.ps]{Two faint, possibly moving objects. The left panels show
pixel maps for the first epoch and the right panels show pixel maps for the second epoch for 
HDF1816 and HDF774. Solid lines cross at the first epoch position. The second epoch position is shown with a box.
Images are 20 pixels on a side (0.6").\label{fig2}}

\figcaption[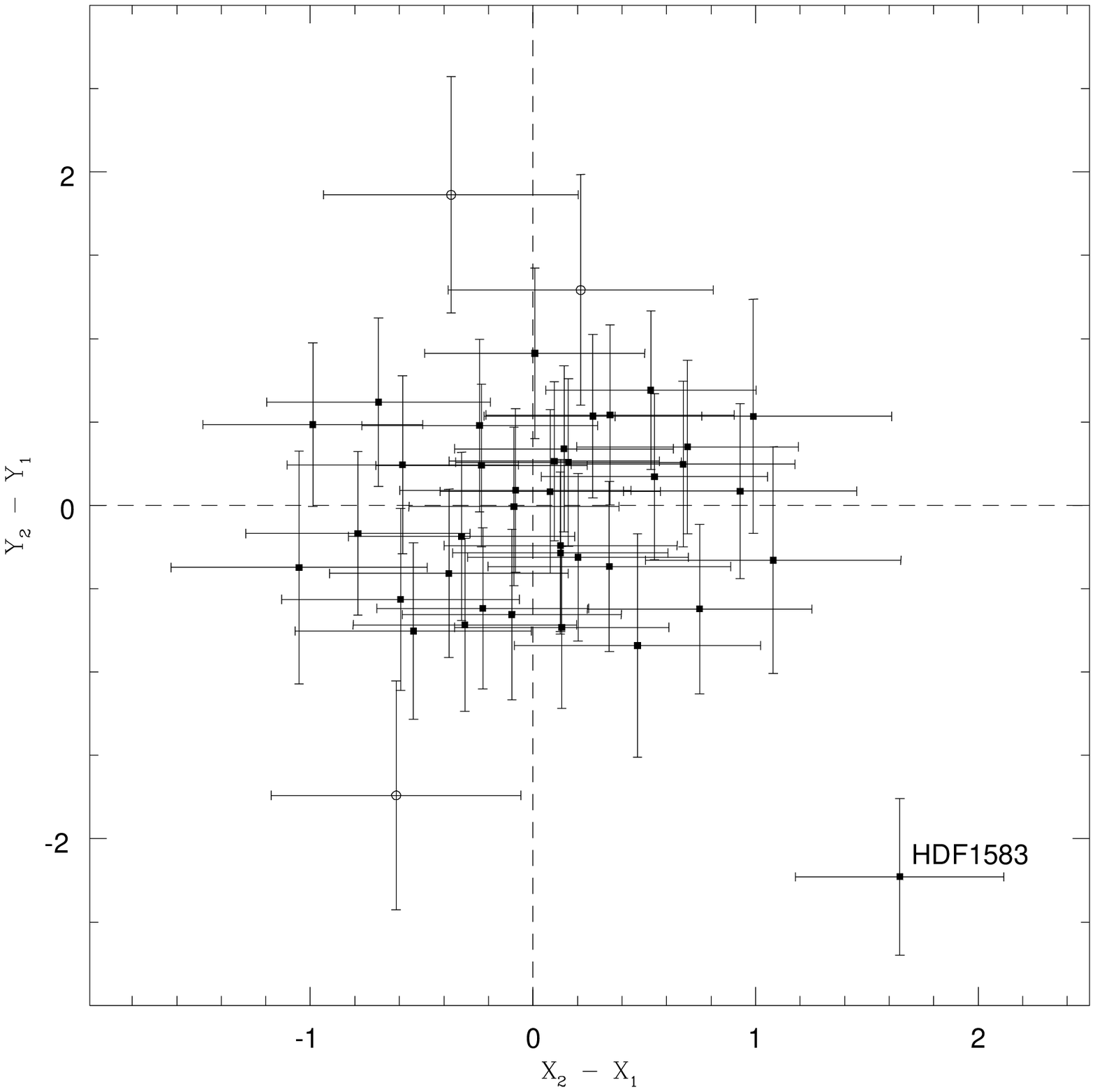]{Difference between second epoch coordinates (X$_2$,Y$_2$)
 and 
transformed first epoch coordinates (X$_1$,Y$_1$) for HDF1583 and the surrounding reference 
compact objects. Reference objects that are not included in our transformations are shown as open circles.
Error bars include centroiding errors from the first and second epochs, and 
the RMS error of the GEOMAP transformation. HDF1583 is an example of an object that is 
clearly exhibiting proper motion.\label{fig3}}

\figcaption[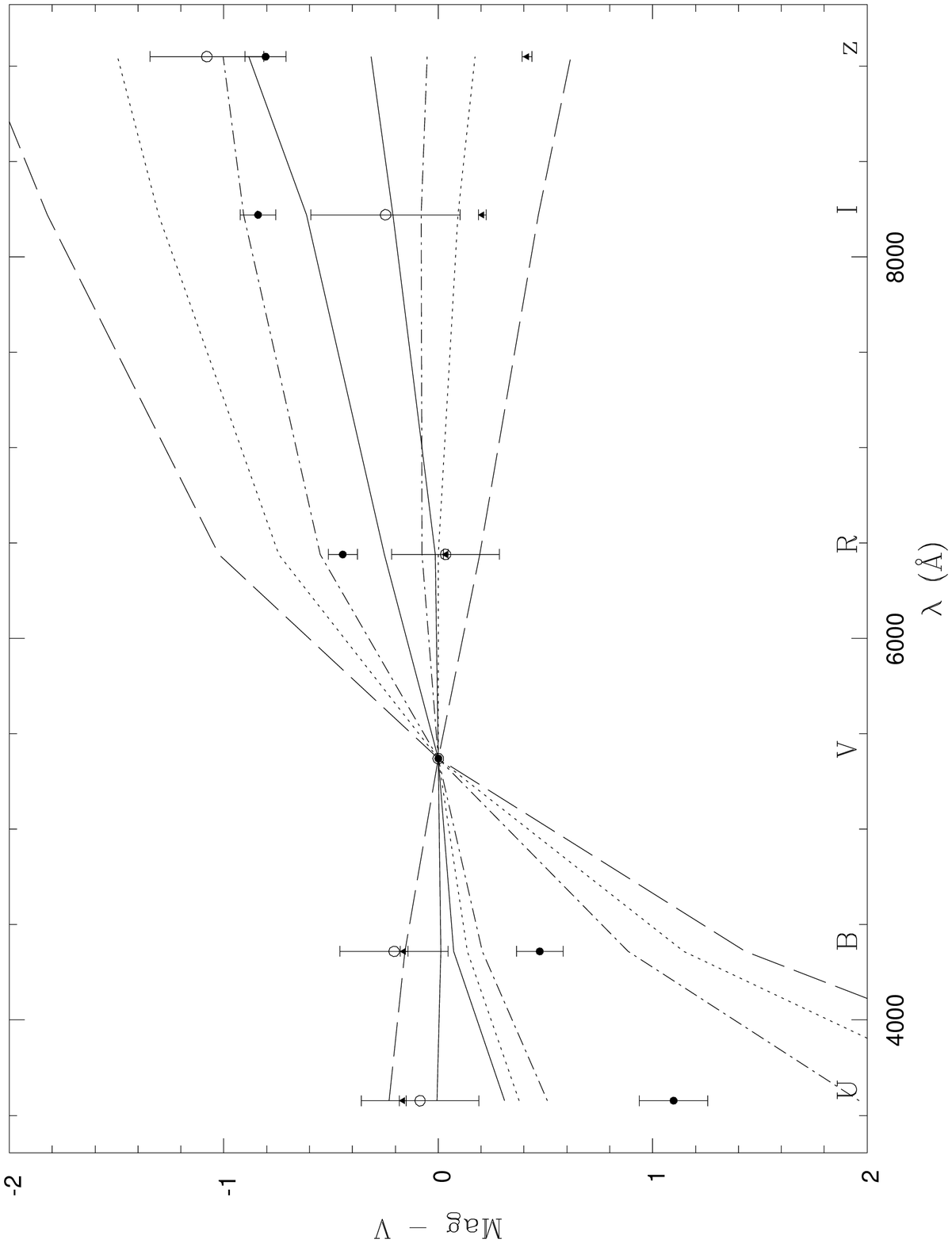]{
Normalized $UBVRIz$ magnitudes
for HDF2234 (filled triangles), HDF3072 (filled circles), and HDF161 (open circles).
The observed magnitudes are normalized at $V$.
The colors for a 15000 and 3000 K blackbody are shown as long--dashed lines. Dashed--dotted
lines represent two DA white dwarf models (7000 K and 3500 K), and dotted lines represent
two DB white dwarf models (8000 K and 3500 K, D. Saumon, private communication). Colors
for a QSO at z=0.3 (upper line) and another at z=0.7 (lower line) are also shown
as solid lines.\label{fig4}}

\figcaption[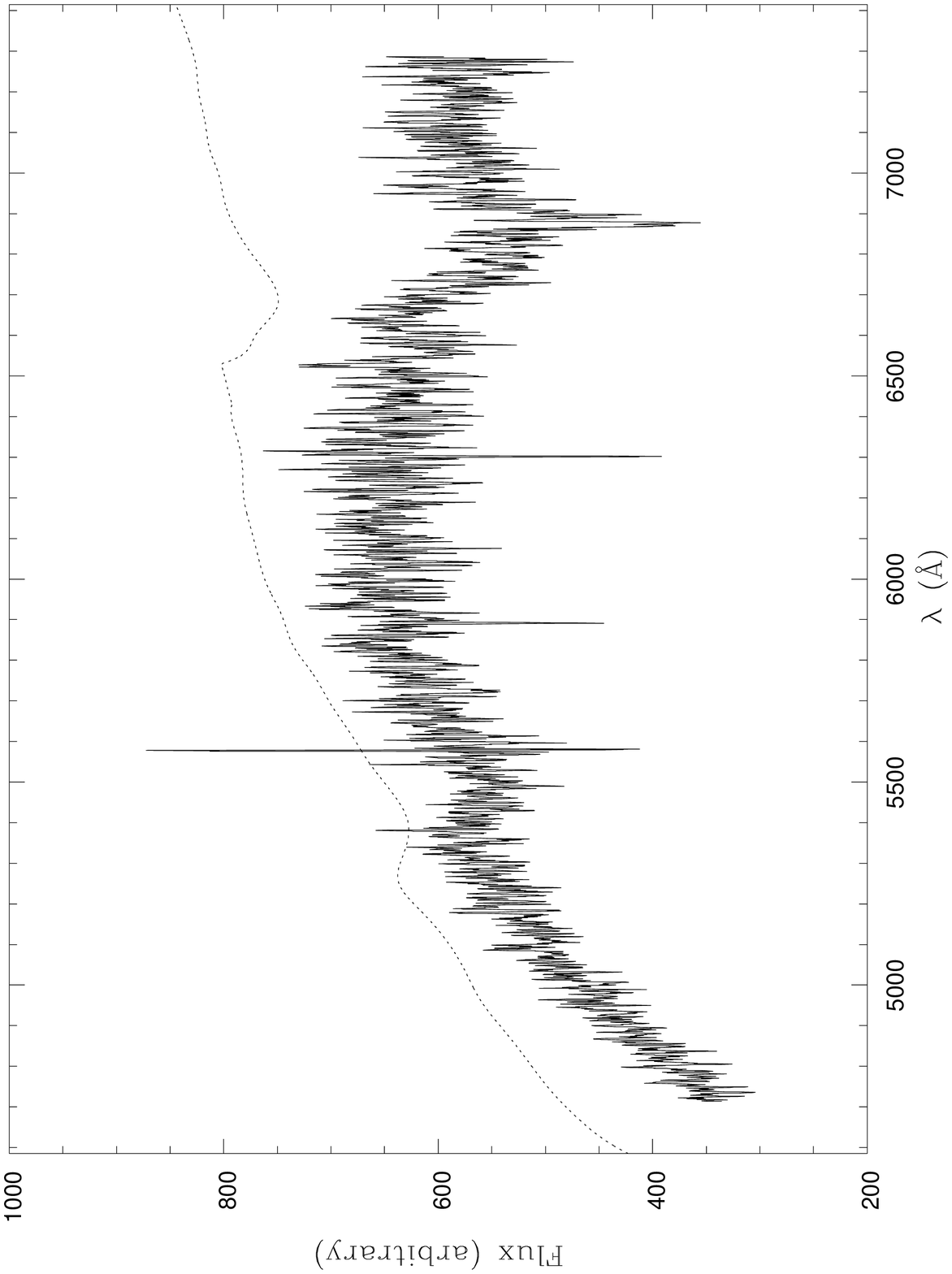]{
Keck/DEIMOS (un--calibrated) spectrum of HDF2234 observed by the Team Keck Treasury Redshift Survey. The spectrum
is smoothed with a 5 pixel wide boxcar.
Instrument efficiency for the blocking filter used for the observations is shown as dotted line.
The broad feature at $\sim$ 5400 \AA\ is due to the instrument efficiency and the feature at
6890 \AA\ is the atmospheric B band. The only detectable feature is H$\alpha$.
\label{fig5}}

\figcaption[f6.ps]{Pixel maps for three of the Ibata et al. (1999) objects, 
HDF2837, HDF2952, and HDF574 with stellarity indices of 0.68, 0.07, and 0.11 respectively. 
Visual inspection of the images further confirms that these are extended objects.
Note that we have not measured proper motions for these objects.\label
{fig6}}

\figcaption[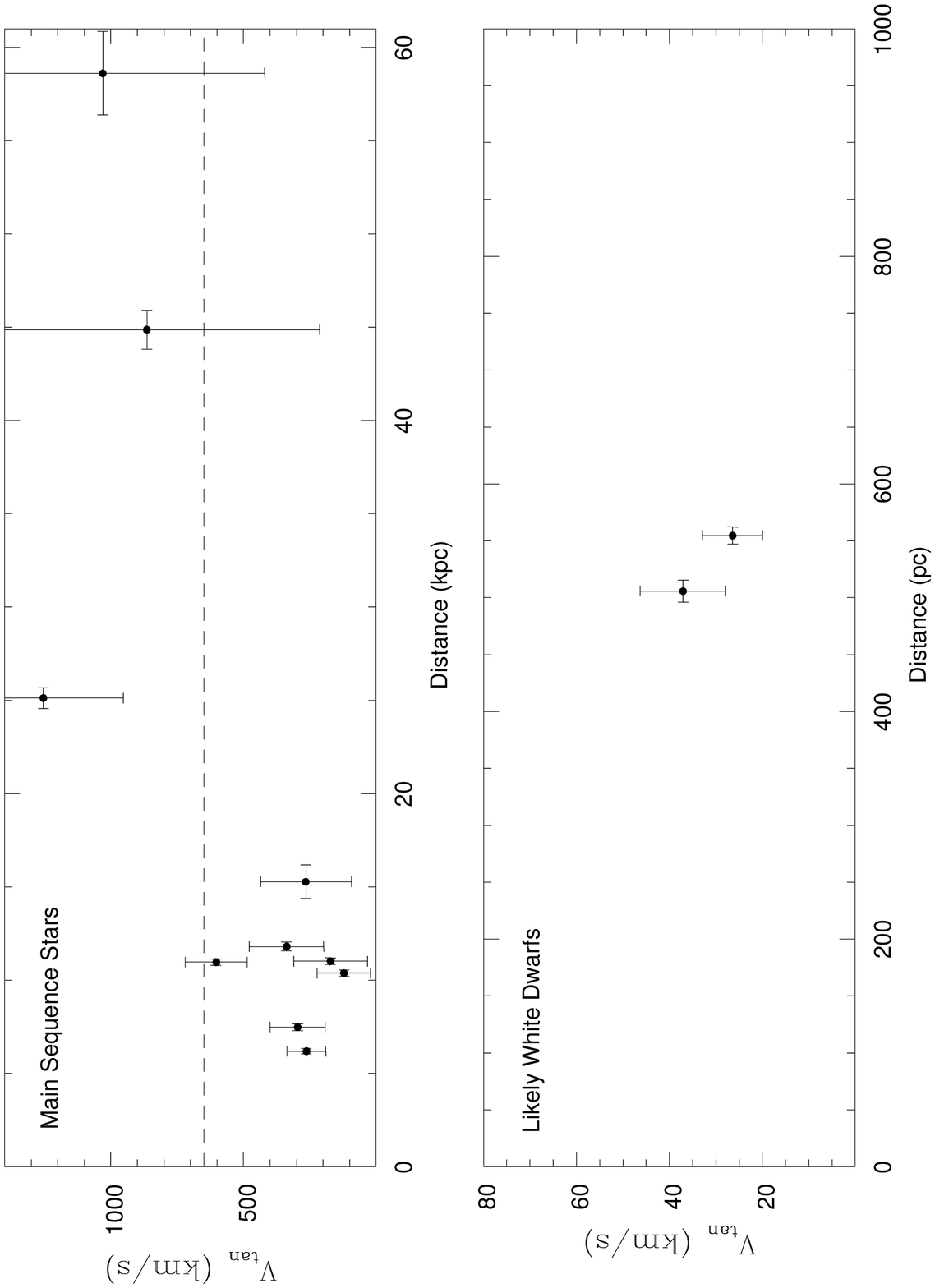]{Tangential velocities and distances for main sequence
stars (top panel) and likely white dwarfs (bottom panel) brighter
than 27th magnitude.
The dashed line marks the upper bound for the escape velocity from 
the Milky Way (650 km sec$^{-1}$; Leonard \& Tremaine 1900; Meillon et al. 1997). All
of the main sequence stars have halo properties, whereas the probable white dwarfs have disk
properties.
\label{fig7}}

\clearpage
\begin{figure}
\plotone{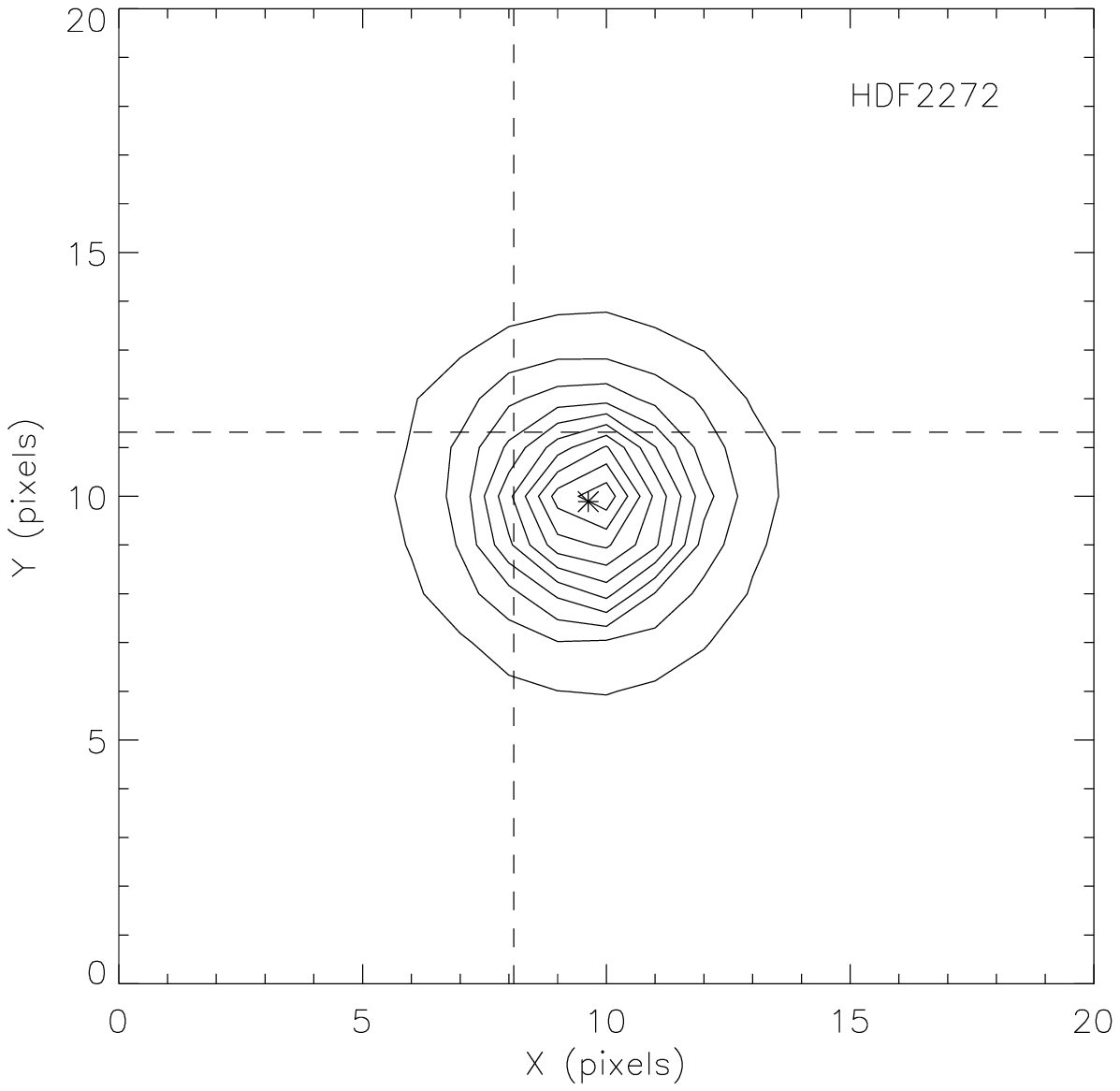}
\plotone{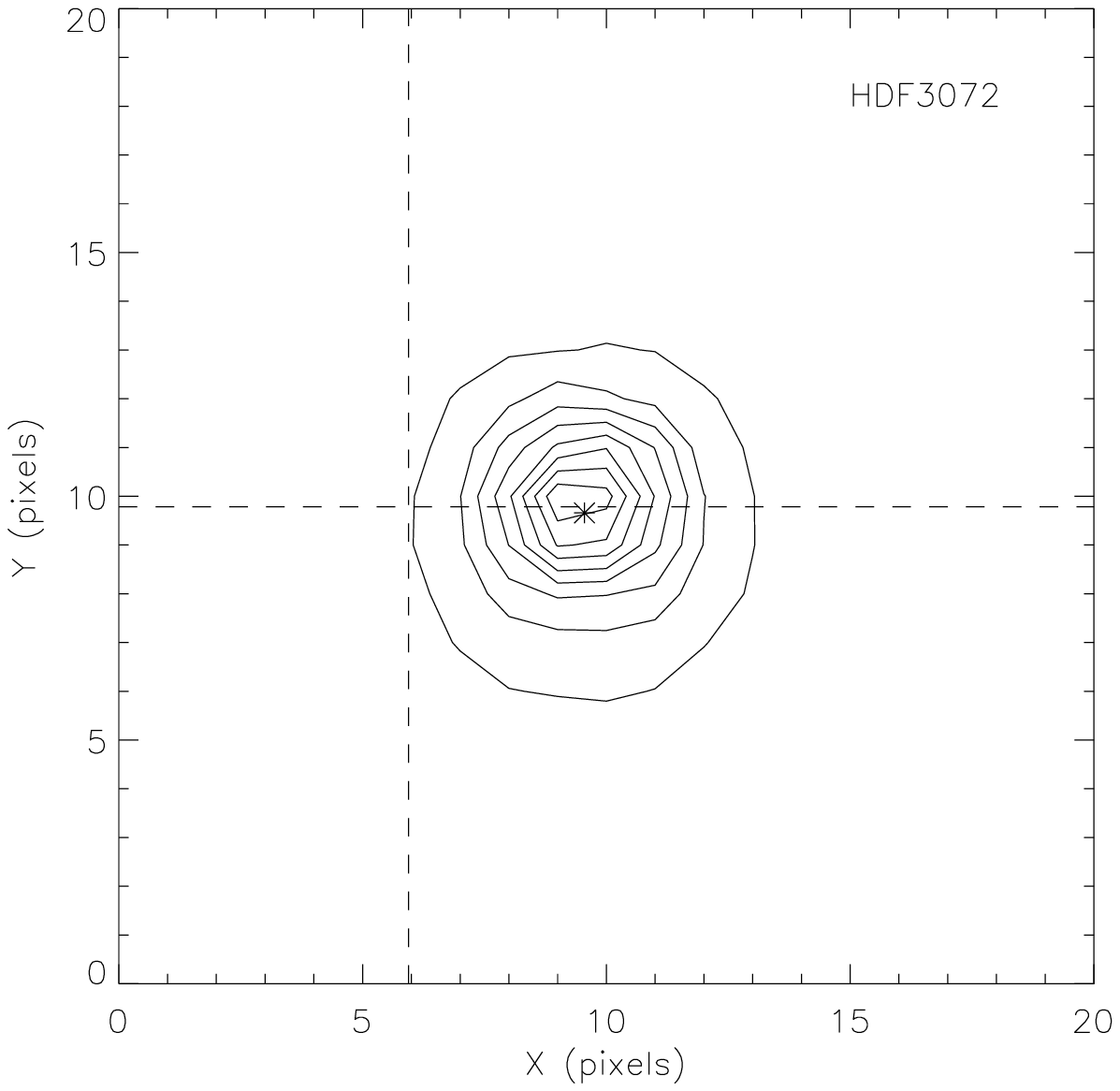}
\begin{flushright}
Figure 1
\end{flushright}
\end{figure}

\clearpage
\begin{figure}
\plottwo{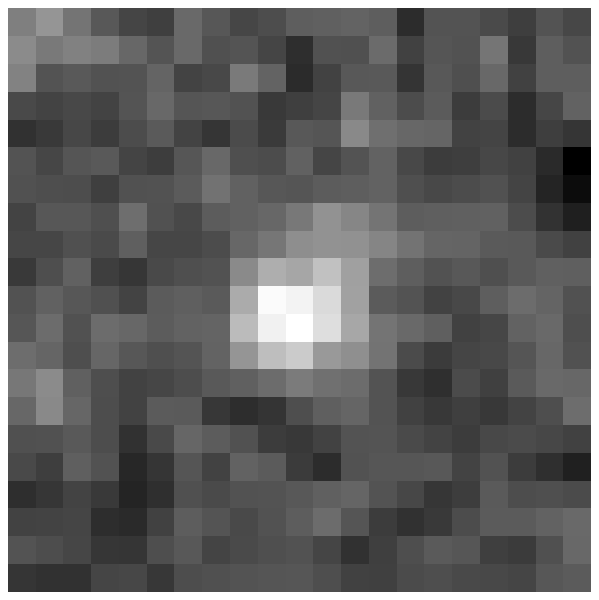}{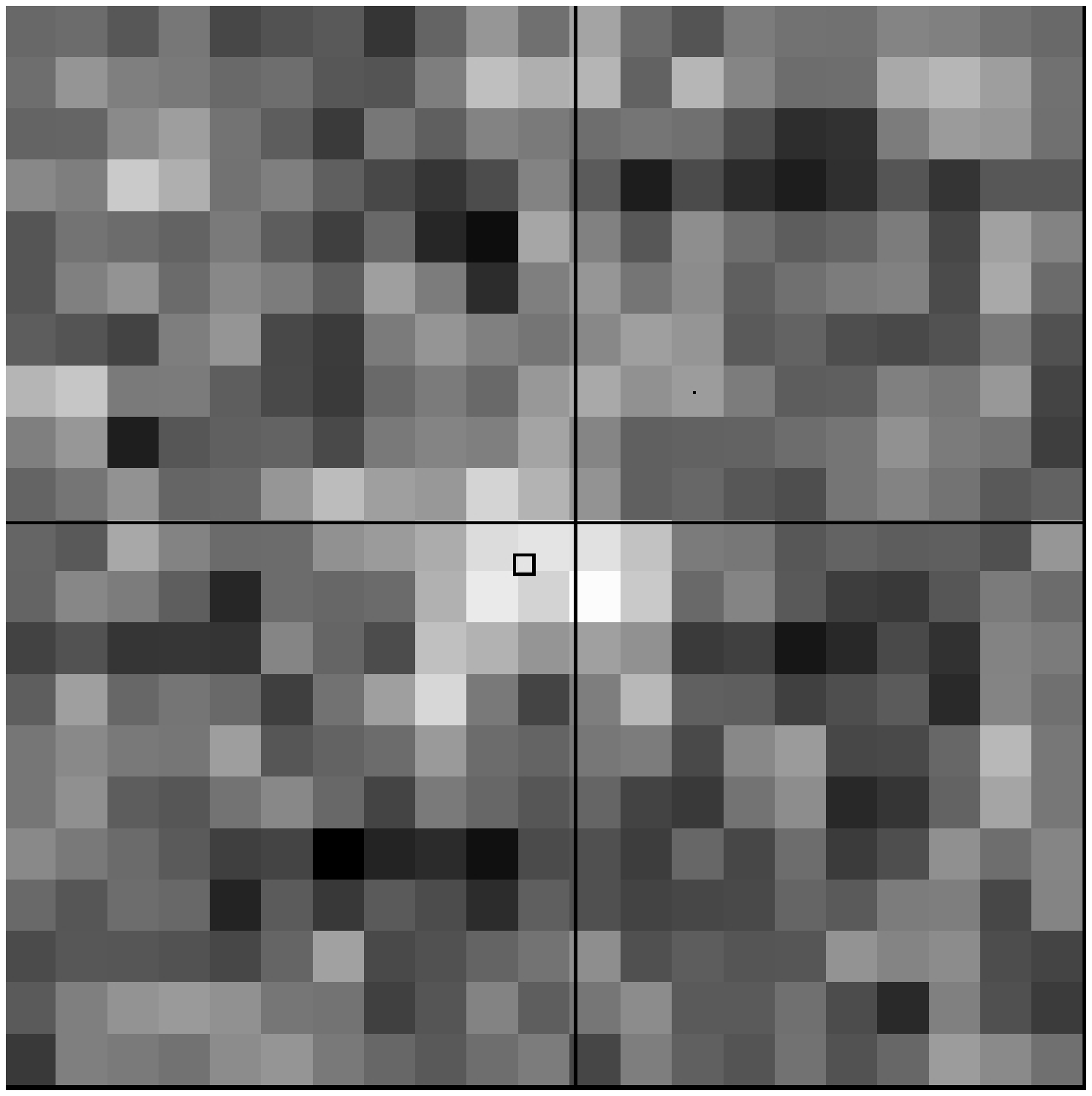}
\end{figure}
\begin{figure}
\plottwo{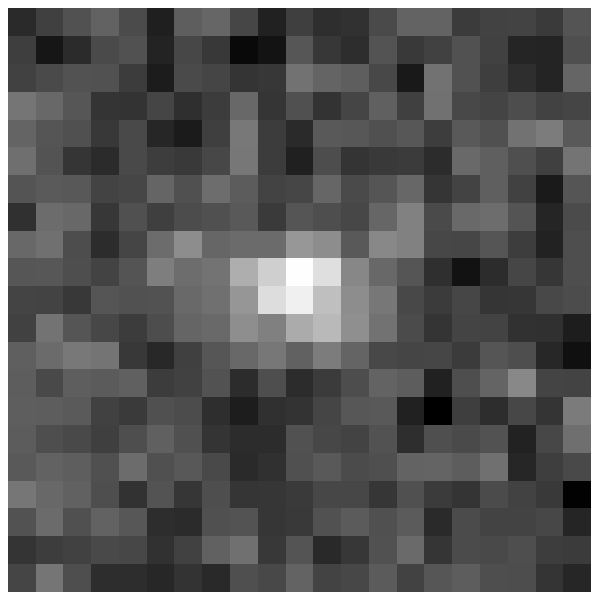}{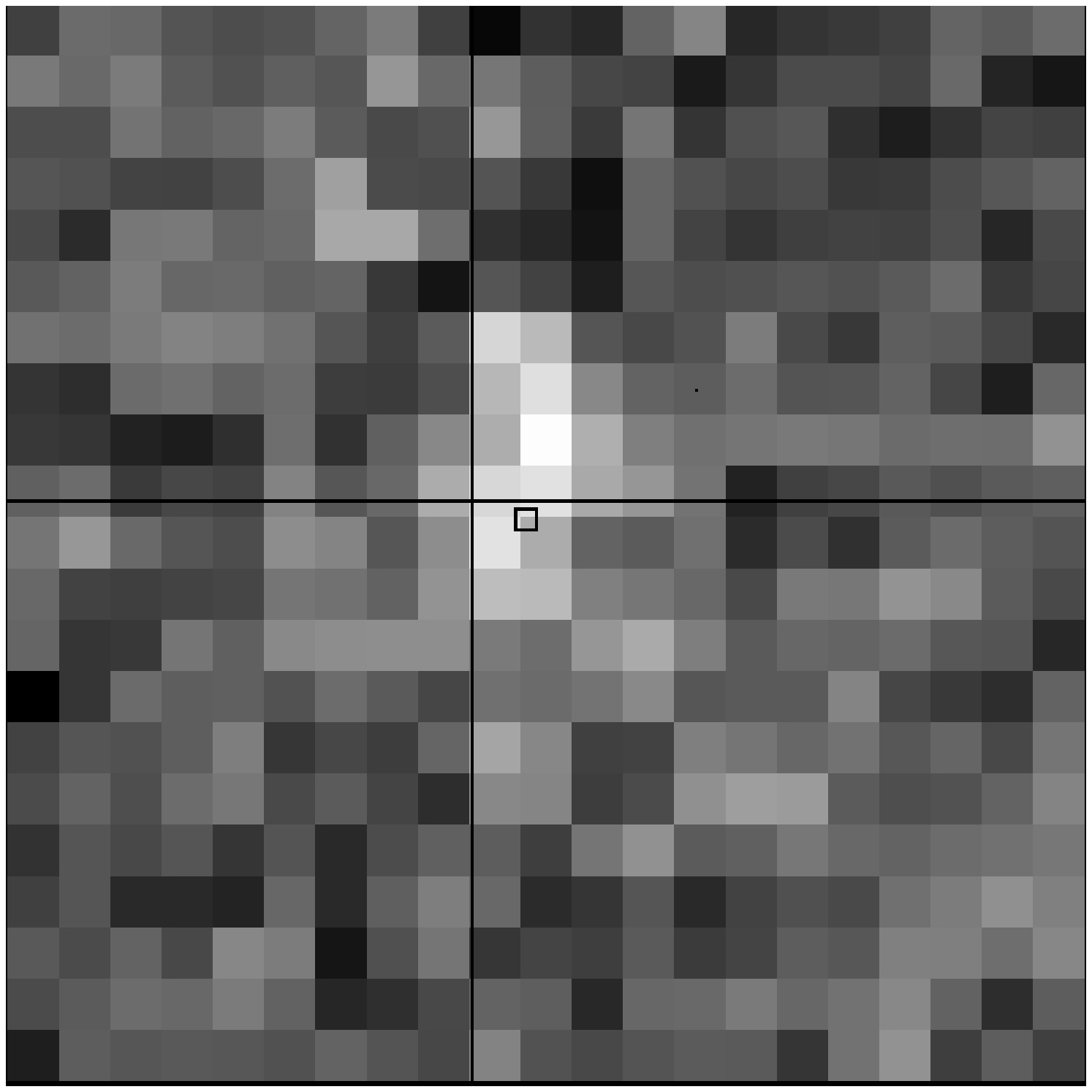}
\begin{flushright}
Figure 2
\end{flushright}
\end{figure}

\clearpage
\begin{figure}
\plotone{f3.ps}
\begin{flushright}
Figure 3
\end{flushright}
\end{figure}

\clearpage
\begin{figure}
\plotone{f4.ps}
\begin{flushright}
Figure 4
\end{flushright}
\end{figure}

\clearpage
\begin{figure}
\plotone{f5.ps}
\begin{flushright}
Figure 5
\end{flushright}
\end{figure}

\clearpage
\begin{figure}
\plottwo{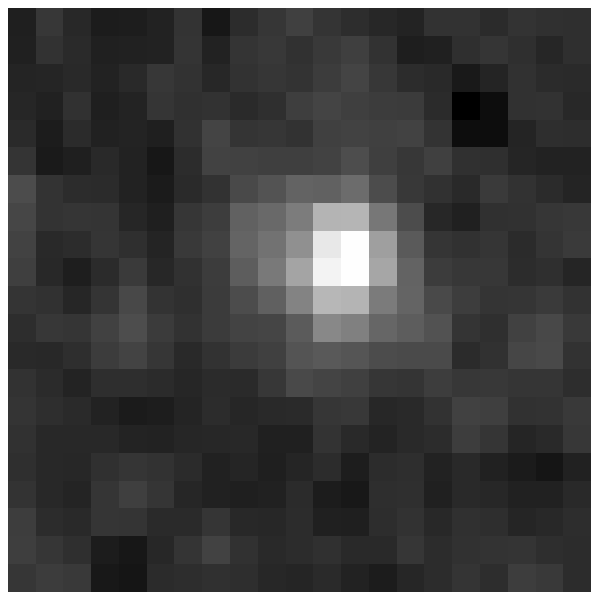}{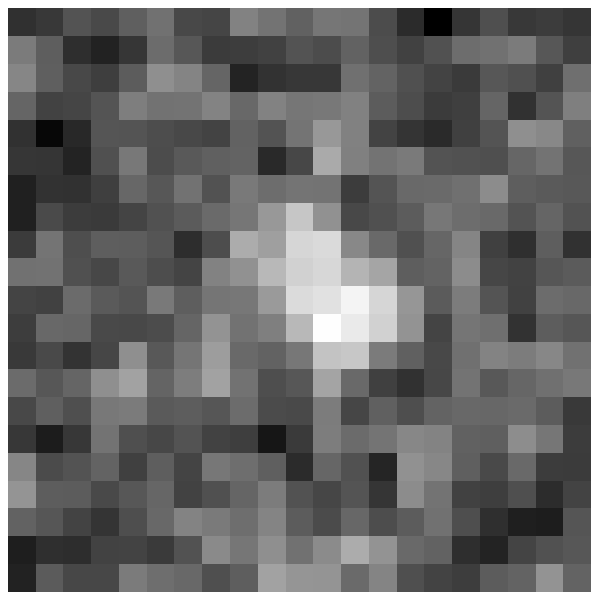}
\end{figure}
\begin{figure}
\plottwo{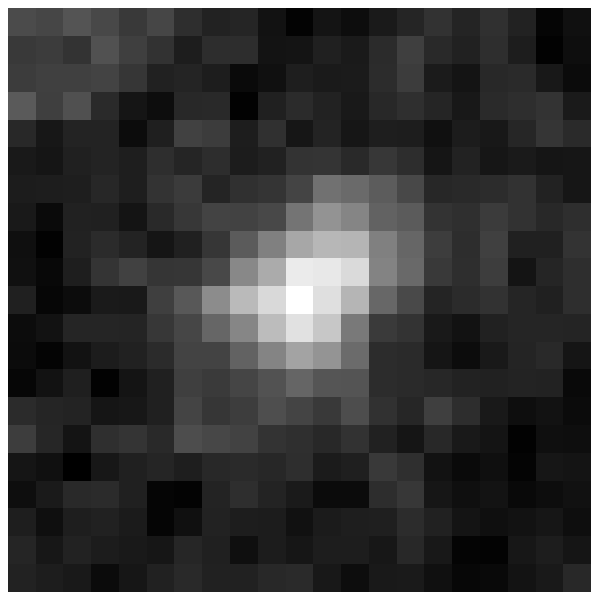}{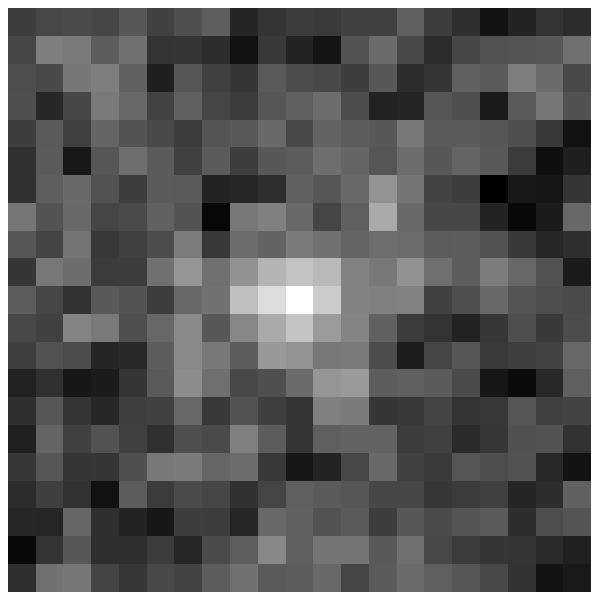}
\end{figure}
\begin{figure}
\plottwo{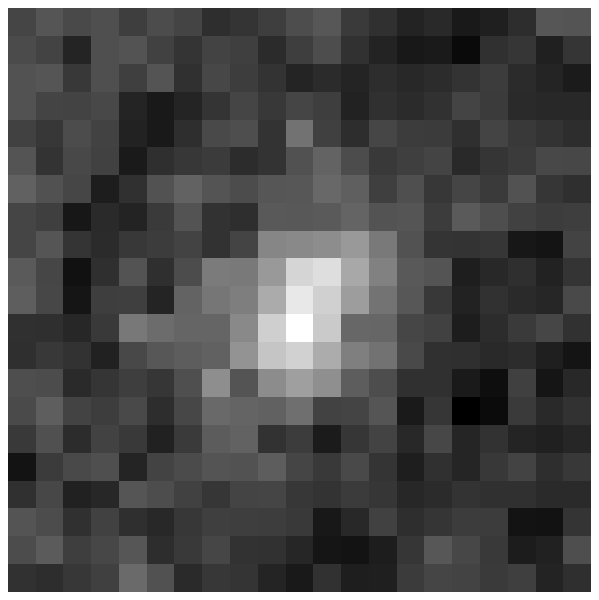}{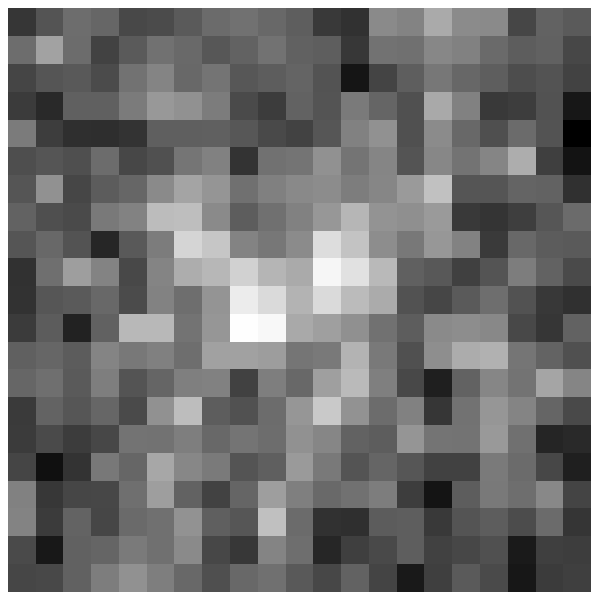}
\begin{flushright}
Figure 6
\end{flushright}
\end{figure}

\clearpage
\begin{figure}
\plotone{f7.ps}
\begin{flushright}
Figure 7
\end{flushright}
\end{figure}


\begin{thebibliography}{}
\bibitem[Afonso et al.(2003)]{afo03} Afonso et al. 2003, \aap, 400, 951
\bibitem[Alcock et al.(1997)]{alc97} Alcock, C. et al. 1997, \apj, 486, 697
\bibitem[Alcock et al.(2000)]{alc00} Alcock, C. et al. 2000, \apj, 542, 281 
\bibitem[Bahcall (1984)]{bah84} Bahcall, J. N. 1984, \apj, 287, 926
\bibitem[Barger et al. (2003)]{bar03} Barger, A. J. et al. 2003, AJ, 126, 632
\bibitem[Beckwith et al.(2003)]{bec03} Beckwith, S. V. W. et al. 2003, BAAS, 202, 1705
\bibitem[Bedin et al.(2003)]{bed03} Bedin, L. R., Piotto, G., King, I. R., and Anderson, J. 2003, \aap, 126, 247
\bibitem[Bergeron(2003)]{ber03} Bergeron, P. 2003, \apj, 586, 201
\bibitem[Bertin \& Arnouts 1996]{ber96} Bertin, E. \& Arnouts, S. 1996, \aaps, 117, 393
\bibitem[Capak et al. (2003)]{cap03} Capak, P. et al. 2003, AJ, in press
\bibitem[Casertano et al.(2000)]{cas00} Casertano, S. et al. 2000, \aap, 120, 2747 
\bibitem[Chabrier et al.(2000)]{cha00a} Chabrier, G., Baraffe, I., Allard, F. \& Hauschildt, P. 2000, \apj, 542, 464
\bibitem[Chabrier et al.(2000)]{cha00b} Chabrier, G., Brassard, P., Fontaine, G. \& Saumon, D. 2000, \apj, 543, 216
\bibitem[Chen et al.(2001)]{che01} Chen, B. et al. 2001, \apj, 553, 184
\bibitem[Cohen et al. (2000)]{coh00} Cohen, J. G. et al. 2000, \apj, 538, 29
\bibitem[Flynn et al.(1996)]{fly96} Flynn, C., Gould, A., and Bahcall, J.N. 1996, \apjl, 466, L55
\bibitem[Giavalisco et al.(2003)]{gia03} Giavalisco, M., and GOODS Team 2003, BAAS, 202, 1703
\bibitem[Giavalisco et al.(2004)]{gia04} Giavalisco, M., and GOODS Team 2004, in preparation
\bibitem[Gilmore et al.(1989)]{gil89} Gilmore, G., King, I. R., and van der Kruit, P.C. 1989, Proceedings of the 19th Advanced Course of the Swiss Society of Astronomy and Astrophysics (SSAA), Saas-Fee, ed. R. Buser, and I. R. King (Mill Valey, University Science Books)  
\bibitem[Hansen et al.(2002)]{han02} Hansen, B. M. S. et al. 2002, \apjl, 574, L155 
\bibitem[Holberg et al.(2002)]{hol02} Holberg, J.B., Oswalt, T. D., and Sion, E. M. 2002, \apj, 571, 512
\bibitem[Ibata et al.(1999)]{iba99} Ibata, R. A., Richer, H. B., Gilliland, R. L., and Scott, D. 1999, \apjl, 524, L95
\bibitem[Ibata et al.(2000)]{iba00} Ibata, R., Irwin, M., Bienaym\'e, O., Scholz, R., and 
	Guibert, J. 2000, \apjl, 532, L41
\bibitem[Kawaler(1996)]{kaw96}Kawaler, S. D. 1996, \apjl, 467, L61
\bibitem[Leonard and Tremaine(1990)]{leo90} Leonard, P. J. T., and Tremaine, S. 1990, \apj, 353, 486
\bibitem[Meillon et al. (1997)]{mei97} Meillon, L., Crifo, F., Gomez, A. E., Udry, S., and 
Mayor, M. 1997, Proceedings of the ESA Symposium 'Hipparcos - Venice '97',
Italy, ESA SP-402 (July 1997), 591
\bibitem[Mendez(2002)]{men02} Mendez, R. A. 2002, \aap, 395, 779
\bibitem[Mendez and Minniti(2000)]{men00} Mendez, R. A., and Minniti, D. 2000, \apj, 529, 911
\bibitem[Oppenheimer et al.(2001)]{opp01} Oppenheimer, B. R., Hambly, N. C., Digby, A. P., Hodgkin, S. T., and Saumon, D. 2001, Science, 292, 698
\bibitem[Pham(1997)]{pha97} Pham, H. A. 1997, ESA SP--402: Hipparcos, Venice, 559
\bibitem[Reid and Majewski(1993)]{rei93} Reid, N., and Majewski, S. R. 1993, \apj, 409, 635
\bibitem[Reid et al.(2001)]{rei01} Reid, I. N., Sahu, K. C., and Hawley, S. L. 2001, \apj, 559, 942
\bibitem[Reyle et al.(2001)]{rey01} Reyle, C., Robin, A. C., and Creze, M. 2001, \aap, 378, L53
\bibitem[Richer et al.(2000)]{ric00} Richer, H. B., Hansen, B. M. S., Limongi, M., Chieffi, A., Straniero, O., and Fahlman, G. G. 2000, \apj, 529, 318
\bibitem[Richer et al.(2001)]{ric01} Richer, H. B. et al. 2001, preprint, astro--ph/0107079
\bibitem[Richer et al.(2002)]{ric02} Richer, H. B. et al. 2002, \apj, 574, 151
\bibitem[Saumon \& Jacobson 1999]{sau99} Saumon, D. \& Jacobson, S.B. 1999, \apj, 511, 107
\bibitem[Udalski et al.(1992)]{uda92} Udalski, A., Szymanski, M., Kaluzny, J., Kubiak, M., 
	and Mateo, M. 1992, AcA, 42, 253
\bibitem[Vanden Berk et al. (2001)]{van01} Vanden Berk D. E. et al. 2001, AJ, 122, 549
\bibitem[Vaucouleurs \& Pence(1978)]{vau78} Vaucouleurs, G. de, \& Pence, W. D. 1978, \aj, 83, 1163
\bibitem[von Hippel \& Bothun]{von90} von Hippel, T. \& Bothun, G. 1990, \aj, 100, 403
\bibitem[Williams et al.(1996)]{wil96} Williams, R. E. et al. 1996, \aj, 112, 1335 
, R. E., Kepler, S. O., \& Lamb, D. Q. 1987, \apj, 315, L77
\bibitem[Wirth et al.(2004)]{wir04} Wirth, G. D. et al. 2004, preprint, astro-ph/0401353
\bibitem[Young (1976)]{you76} Young, P. J. 1976, \aj, 81, 807
\end{thebibliography}
\end{document}